\documentclass[conference]{IEEEtran}
\IEEEoverridecommandlockouts

\usepackage{cite}
\usepackage{amsmath,amssymb,amsfonts}
\usepackage{algorithmic}
\usepackage{graphicx}
\usepackage{textcomp}
\usepackage{xcolor}
\usepackage{url}
\usepackage{booktabs}
\usepackage{float}

\def\BibTeX{{\rm B\kern-.05em{\sc i\kern-.025em b}\kern-.08em
    T\kern-.1667em\lower.7ex\hbox{E}\kern-.125emX}}
\begin{document}

\title{Rashomon perspective for measuring uncertainty in the survival predictive maintenance models* \thanks{*Presented at 33. IEEE Conference on Signal Processing and Communications Applications}}

\author{\IEEEauthorblockN{Yigitcan Yardimci}
\IEEEauthorblockA{\textit{Department of Statistics} \\
\textit{Eskisehir Technical University}\\
Eskisehir, Turkiye \\
yigitcanyardimcii@gmail.com}
\and
\IEEEauthorblockN{Mustafa Cavus}
\IEEEauthorblockA{\textit{Department of Statistics} \\
\textit{Eskisehir Technical University}\\
Eskisehir, Turkiye \\
mustafacavus@eskisehir.edu.tr}
}

\maketitle

\begin{abstract}


The prediction of the Remaining Useful Life of aircraft engines is a critical area in high-reliability sectors such as aerospace and defense. Early failure predictions help ensure operational continuity, reduce maintenance costs, and prevent unexpected failures. Traditional regression models struggle with censored data, which can lead to biased predictions. Survival models, on the other hand, effectively handle censored data, improving predictive accuracy in maintenance processes. This paper introduces a novel approach based on the Rashomon perspective, which considers multiple models that achieve similar performance rather than relying on a single best model. This enables uncertainty quantification in survival probability predictions and enhances decision-making in predictive maintenance. The Rashomon survival curve was introduced to represent the range of survival probability estimates, providing insights into model agreement and uncertainty over time. The results on the CMAPSS dataset demonstrate that relying solely on a single model for RUL estimation may increase risk in some scenarios. The censoring levels significantly impact prediction uncertainty, with longer censoring times leading to greater variability in survival probabilities. These findings underscore the importance of incorporating model multiplicity in predictive maintenance frameworks to achieve more reliable and robust failure predictions. This paper contributes to uncertainty quantification in RUL prediction and highlights the Rashomon perspective as a powerful tool for predictive modeling. 

\end{abstract}


\section{Introduction} 

Equipment failures pose significant challenges in the industry \cite{kane_et_al_2022}. Minimizing downtime from mechanical failures is critical for operational continuity. Predictive maintenance (PM) identifies potential failures and optimizes maintenance schedules to reduce costs and disruptions \cite{castle_et_al_2020}. Industry 4.0-based PM uses real-time system monitoring to detect failures early \cite{cummins_et_al_2024}, enhancing operational efficiency \cite{hasib_et_al_2023}. AI-driven methods further reduce costs through anomaly detection, fault diagnosis, and Remaining Useful Life (RUL) estimation. It is the process of predicting the remaining operational time or lifespan of a system or component before it fails. It helps in scheduling maintenance and preventing unexpected breakdowns \cite{pashami_et_al_2023}. 

The literature on the RUL estimation is dominated by regression modeling. However, the existence of censored data ---an event of interest (e.g., failure) that has not occurred by the end of the study period or is only known to occur within a certain time interval--- undermines the superiority of regression models. Fortunately, survival models can handle the censored data and thus are not affected by survival bias.

Survival models, used for RUL estimation, assess failure probability over time and aid in efficient maintenance planning. Proven effective in predicting system evolution \cite{holmer_et_al_2023}, it is widely used in medicine and increasingly in engineering for failure time prediction. It provides essential information on periods without failure and must be incorporated to avoid analysis errors \cite{zeng_et_al_2023, Alabdallah_2025}. We apply survival analysis models for aircraft engine RUL estimation on the CMAPSS dataset \cite{saxena_et_al_2008} for considering censored data. However, relying on a single model in such predictions, i.e. ignoring other good models, may lead to increased prediction error at certain time points and therefore unnecessary repair costs or serious problems that may arise from maintenance not performed although necessary because of uncertainty in the predictions. On the other hand, uncertainty quantification is another significant challenge in such problems \cite{vollert_and_theissler_2021}. In this regard, we propose a new approach called \textit{Rashomon survival curve} based on the Rashomon perspective \cite{rudin_et_al_2024} which is derived from Rashomon effect \cite{breiman_2001} ---considering multiple models that achieve similar performance levels based on different modeling assumptions on the same dataset instead of relying on the best model--- to predict survival probability. It also enables quantifying uncertainty of the survival probability predictions within the intended time interval. Our approach is different from the classical ensembling approach \cite{madras_et_al_2019}, as it considers \textit{multiple high-performing models rather than relying solely on the ensemble of models}. To the best of our knowledge, this is the first paper to consider the Rashomon perspective as a solution in the survival models. From this perspective, this paper introduces an important phenomenon with the potential to provide solutions to various issues in predictive modeling. Moreover, we investigate the effect of different censoring times ($t = 200$, $225$, $250$) on the prediction uncertainty in the CMAPSS dataset. Our findings contributed to showing the potential of the Rashomon perspective in survival models, quantifying the prediction uncertainty of survival models, and understanding the impact of censoring times on survival probability prediction and uncertainties in the modeling process.


\section{Methodology} 
This section defines the concepts of the Rashomon survival set and the Rashomon survival curve. 

\subsection{Rashomon Set}

The Rashomon set is defined as the set of all predictive models within a given hypothesis space $\mathcal{F}$, and loss function $L$ that is close to the best-performing model in terms of prediction performance \cite{marx_et_al_2020}:
\begin{equation}
    \hat{R}(\mathcal{F}, \varepsilon) = \{ f \in \mathcal{F} : \hat{L}(f) \leq \hat{L}(f^*) + \varepsilon \}    
\end{equation}

\noindent where $\hat{L}(f)$ is empirical risk, $f^*$ is the model with the lowest error, and $\varepsilon$ is the acceptable error margin (aka. Rashomon parameter).  

\subsection{Rashomon Survival Set}
The Rashomon survival set is defined as the set of survival models within a given hypothesis space $F_S$, and loss function $L$ that yield predictions close to the most accurate survival model in terms of survival time estimation. The Rashomon set for survival models is expressed as follows:

\begin{equation}
    \hat{R}_S(\mathcal{F_S}, \varepsilon) = \{ f_s \in \mathcal{F_S} : \hat{L}(f_s) \leq \hat{L}(f^*_s) + \varepsilon \}
\end{equation}

\noindent where $\hat{L}(f_s)$ is the calculated error value of the survival model (e.g., c-index, Brier score),  $f^*_S$ is the most accurate survival model, and $\varepsilon$ is the acceptable error margin.

\subsection{Rashomon Survival Curve}
The Rashomon survival curve ($\Delta_{\epsilon}$) is defined as representing the lowest and highest survival estimation $S(t_i)$ for time $t_i$, similar to the \textit{viable prediction range} \cite{watson_et_al_2023}, by the models in the Rashomon survival set:

\begin{equation}
    \Delta_{\epsilon}(t_i) = 
    \left[ 
    \min_{s \in R_s} S(t_i), 
    \max_{s \in R_s} S(t_i) 
    \right]
\end{equation}

\noindent It measures the uncertainty in the survival prediction of the models in the Rashomon survival set. By examining the survival prediction ranges over time for a given individual, we can assess how different models diverge in their estimation of survival probabilities at various time points. This approach highlights the degree of temporal uncertainty or consensus in model predictions, enabling a deeper understanding of how survival risks evolve and identifying cases where model disagreement is particularly pronounced at specific time intervals.

\section{Experiments} 

In this paper, survival models are used to predict the RUL of the engines. The process consists of data pre-processing and partitioning, model training, censored data strategy, performance evaluation, and visualization of the results. The materials for reproducing the experiments and the datasets can be found in the repository: \url{https://github.com/mcavs/Rashomon_survival_curve_paper}

\subsection{Dataset}

In this paper, the CMAPSS dataset is used to analyze the performance of jet engines and predict the RUL of the engines. It is designed to model jet engine performance under various operating conditions and fault modes consisting of four main subsets (\texttt{FD001, FD002, FD003, FD004}), each representing different operational scenarios, failure modes, and sample sizes $100$, $260$, $100$, and $246$, respectively. Table~\ref{tab:cmapss_detailed} summarizes the variables in the CMAPSS dataset. During the preprocessing phase, variables that remained constant across all cycles were removed to eliminate redundancy and improve model performance. 



\begin{table*}[h!]
\centering
\caption{\textbf{Detailed Summary of Variables in the CMAPSS Dataset }}
\label{tab:cmapss_detailed}
\setlength{\tabcolsep}{5pt} 
\renewcommand{\arraystretch}{1.1} 
\begin{tabular}{p{2.5cm} p{3.5cm} p{2.5cm} p{2.5cm} p{2.5cm} p{2.5cm}}\toprule
&& \multicolumn{4}{c}{\textbf{Range/Values}} \\\cline{3-6}
                  \textbf{Variable} & \textbf{Description} & \texttt{FD001} &  \texttt{FD002} & \texttt{FD003} &  \texttt{FD004} \\\midrule
\texttt{unit\_number} & Unique identifier for each engine & \{1, 2, ..., 100\} & \{1, 2, ..., 260\} & \{1, 2, ..., 100\} & \{1, 2, ..., 260\}\\
\texttt{time\_in\_cycles} & Time in operational cycles since engine start & \{1, 2, ..., 362\} & \{1, 2, ..., 378\} & \{1, 2, ..., 525\} & \{1, 2, ..., 378\}\\
\texttt{op\_set\_1} & Operational condition parameter 1 & $[-0.0087, 0.0087]$ & $[0.000, 42.008]$ & $[-0.0086, 0.0086]$ & $[0.000, 42.008]$\\
\texttt{op\_set\_2} & Operational condition parameter 2 & $[-6e-04, 6e-04]$ & $[0.000, 0.842]$ & $[-6e-04, 7e-04]$ & $[0.000, 0.842]$\\
\texttt{op\_set\_3} & Operational condition parameter 3 & $-$ & $[60, 100]$ & $-$ & $[60, 100]$\\
\texttt{sensor\_1} & Measurement from sensor 1 & $-$ & $[445.00, 518.67]$ & $-$ & $[445.00, 518.67]$\\
\texttt{sensor\_2} & Measurement from sensor 2  & $[641.21, 644.53]$ & $[535.53, 644.52]$ & $[640.84, 645.11]$ & $[535.53, 644.52]$\\
... & ... & ... & ... & ... & ...\\
\texttt{sensor\_21} & Measurement from sensor 21 & $[22.8942, 23.6184]$ & $[6.0105, 23.5901]$ & $[22.8726, 23.9505]$ & $[6.0105, 23.5901]$\\\bottomrule
\end{tabular}
\begin{flushleft}
\footnotesize 
$-$ These variables were removed from FD001 and FD003 due to being constant (i.e., non-informative).
\end{flushleft}
\end{table*}

\subsection{Modeling}

$19$ different survival models were used: Cox, Gradient Boosting Machine \cite{buhlmann2007}, Accelerated Oblique Random Survival Forest \cite{jaeger_et_al_2019}, BlackBoost, CForest \cite{hothorn2004}, CoxBoost \cite{binder2008coxboost}, CTree \cite{hothorn2006ctree}, Cross-Validated CoxBoost \cite{binder2008coxboost}, Cross-Validated GLMNet \cite{friedman2010cvglmnet}, GLMBoost \cite{hofner2014model}, GLMNet, Kaplan-Meier, Nelson-Aalen \cite{aalen1978nonparametric}, Penalized Cox\cite{royston2002}, Random Forest (Ranger) \cite{wright2017}, Random Survival Forest (SRC) \cite{ishwaran2008}, RPart \cite{therneau1997rpart}, XGBoost AFT \cite{kvamme2019}, and XGBoost Cox \cite{kvamme2019}. Each dataset was used as $80\%$ training and $20\%$ testing for modeling. The censoring times are fixed as $t = 200, 225, 250$ to enable predictions at different time stages regarding the RUL of the engines. Then, the value of $\varepsilon = 0.05$ is selected for the creation of the Rashomon survival set, similar to studies in the literature \cite{muller_et_al_2023, cavus_and_biecek_2024}.

\section{Results} 

The Rashomon survival sets were evaluated using the concordance index (C-index). For \texttt{FD001}, 5 models in the Rashomon survival set with the mean and standard deviation of the C-index of $0.8259 \pm 0.0204$, respectively. Similarly, \texttt{FD002} included 4 models ($0.7189 \pm 0.0124$), \texttt{FD003} consisted of 4 models ($0.8707 \pm 0.0181$), and \texttt{FD004} contained 8 models ($0.8027 \pm 0.0146$). The performance of the sets shows we achieved the highest-performant models for \texttt{FD003}, and the lowest one for \texttt{FD002}. The other sets with relatively similar performance around $0.8$.

\begin{figure*}
    \centering
    \caption{The survival probability curve of the data subsets \texttt{FD001}, \texttt{FD002}, \texttt{FD003}, \texttt{FD004} for the censoring times $t = 200, 225, 250$. The black stepwise line shows the Rashomon survival curve, and the gray band around it represents the survival prediction range of the models in the Rashomon set.}
    \label{fig:facet_plot}
    \includegraphics[width=\linewidth]{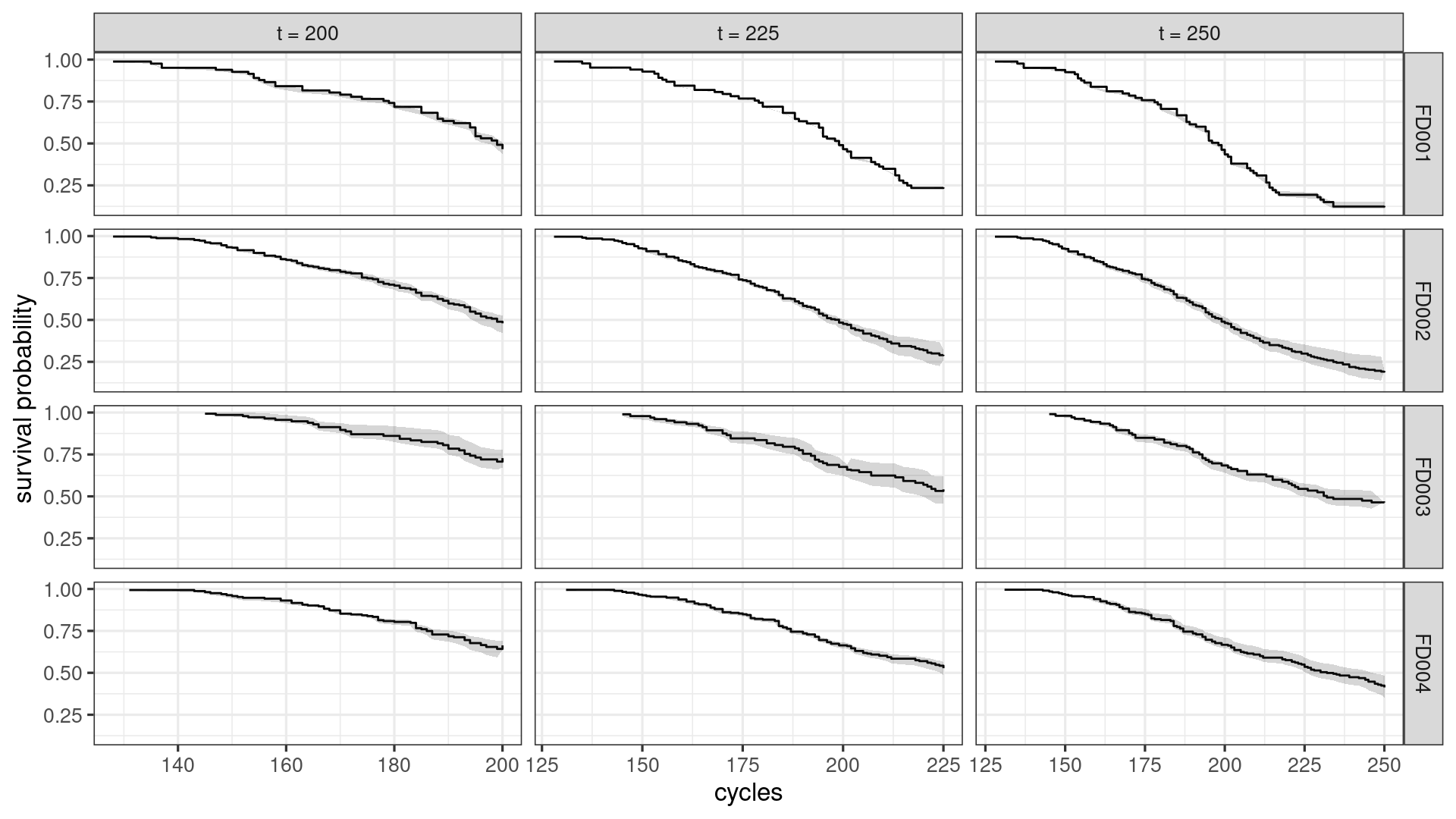}
\end{figure*}

Figure~\ref{fig:facet_plot} illustrates the survival probability curves and uncertainty of four different subsets at three censoring times $t = 200, 225, 250$. The surrounding shaded region represents the prediction uncertainty of the survival probability calculated on the Rashomon survival set. A wider shaded region suggests greater variability or uncertainty in the survival probability, whereas a narrower band indicates a more precise estimation. 

Survival probability varies significantly across different subsets. Subsets like \texttt{FD001}, which exhibit higher survival probabilities (above $75\%$ at $250$ cycles), represent more durable systems in the long run, while subsets like \texttt{FD002}, where the survival curves decline more rapidly (below $50\%$ at $200$ cycles and near $25\%$ at $250$ cycles), indicate shorter-lived systems. As the censoring time increases, overall survival probabilities tend to decrease, providing a clearer picture of long-term reliability. In particular, \texttt{FD002} and \texttt{FD003} show lower survival rates over extended cycles (below $50\%$ at $225$ cycles), whereas \texttt{FD001} maintains relatively higher survival probabilities (over $75\%$ at $250$ cycles) for a longer duration. This suggests that different operational conditions or failure mechanisms influence the longevity and reliability of each subset.

Similarly, uncertainty levels vary across different subsets and censoring times. Subsets like \texttt{FD001} have the narrowest confidence intervals (below $±5\%$ at $250$ cycles), indicating more consistent failure patterns and predictable degradation. In contrast, \texttt{FD002} exhibits the highest uncertainty (exceeding $±15\%$ at $250$ cycles), suggesting a diverse range of failure times and greater variability in operational conditions. As censoring time increases, uncertainty grows due to fewer remaining observations, particularly in \texttt{FD002} and \texttt{FD003}, where confidence intervals exceed $±12\%$ at $225$ cycles and $±15\%$ at $250$ cycles. Meanwhile, \texttt{FD004} shows moderate uncertainty (around $±8\%$ at $250$ cycles), and \texttt{FD001} remains the most predictable subset. These variations suggest that different subsets experience different levels of failure consistency, impacting the reliability of survival probability estimates over time.

\section{Conclusions} 

We defined a new approach based on the Rashomon perspective for measuring prediction uncertainty in survival predictive maintenance models. Accordingly, using the CMAPSS dataset, 19 different survival models were trained, and model performance was analyzed using evaluation metrics then the Rashomon survival sets were created. 

The results highlight that relying on a single model for RUL estimation may be a higher risk in some subsets. Moreover, uncertainty levels of survival predictions vary across different censoring times and subsets. Uncertainty increases due to fewer remaining observations while censoring time increases. These results indicate that our approach is an effective method for quantifying uncertainty in RUL prediction. However, these findings are limited to the CMAPSS dataset. 

In conclusion, these findings demonstrate the role of our approach specifically in RUL estimation for predictive maintenance scenarios. In addition, it is seen that the Rashomon perspective is an important solution tool for basic modeling problems in general. As further work, conducting large-scale studies that consider different datasets and modeling approaches will be crucial in assessing the generalizability of this method.

\section*{Acknowledgment} 
The work on this paper is supported by the Artificial Intelligence Talent Cluster of Defence Industry Academic Thesis Program (SAYZEK - ATP) organized in collaboration with the Presidency of the Republic of Turkiye - Secreterial of Defense Industries and the Council of Higher Education.


\end{document}